\documentclass[12pt]{article}
\usepackage{epsfig}
\usepackage{latexsym}
\begin{document}
\title{Renormalization by enforcing a symmetry.}
\author{ A.A.Slavnov \thanks{E-mail:$~~$ slavnov@mi.ras.ru}
\\Steklov Mathematical Institute, Russian Academy 
of Sciences\\ Gubkina st.8, GSP-1,117966, Moscow\\ and \\ Max-Planck-
Institute f\"{u}r Physik (Werner-Heisenberg-Institute)\\ F\"{o}hringer
Ring 6, 80805 M\"{u}nchen} \maketitle

\begin{abstract}
A new renormalization scheme for theories with nontrivial internal 
symmetry is proposed. The sheme is regularization independent and 
respects the symmetry requirements.
\end{abstract}

\section {Introduction}

Renormalization procedure in the theories with nontrivial internal 
symmetries, like gauge invariant models, is complicated by the necessity 
to provide the symmetry of the renormalized theory. The crucial role in 
this procedure is played by the relations between Green functions which 
are the quantum analogue of a classical symmetry. In the case of Quantum 
Electrodynamics (QED) these relations are Ward-Takahashi identities (WTI) 
\cite{W}, \cite{T}, which connect three- and two-point Green functions. 
For non-Abelian gauge theories corresponding relations (STI)were obtained 
in the papers \cite{Sl}, \cite{Tl}.

There are essentially two approaches to renormalization of gauge 
invariant theories. The first one uses some intermediate gauge invariant 
regularization, e.g. dimensional regularization \cite{H.V}, \cite{B.M}, 
higher covariant derivatives \cite{Sl2}, \cite{L.Z}, \cite{B.S}, or 
lattice regularization \cite{Wil}. Using these regularizations one can 
prove that the counterterms needed to eliminate ultraviolet divergencies 
preserve the gauge invariant structure of the renormalized Lagrangian.

In the second approach, known as algebraic renormalization, one firstly 
defines a finite renormalized theory following the Bogoliubov-Parasiuk 
R-operation \cite{B.P} with some particular subtractions \cite{Z}. This 
procedure in general breaks gauge invariance, however one can use a 
finite renormalization freedom to restore STI for renormalized Green 
functions \cite{BRS}, \cite{P.R} (for recent development and more 
complete references see \cite{Gr}).

Both these approaches have some advantages and disadvantages. Using 
invariant regularizations allows to prove in a rather simple way gauge 
independence of observables. Moreover the dimensional regularization 
proved to be an efficient way for calculations of Feynman diagrams. On 
the other hand in this case the procedure is explicitely regularization 
dependent and, more important, is not directly applicable to some 
theories including the Standard Model and supersymmetric theories. The 
invariant regularization procedure for anomaly free theories including 
chiral fermions, like Salam-Weinberg or unified models was constructed 
recently \cite{F.S}, \cite{D.S}, and for general Feynman diagrams is 
still rather complicated.

The algebraic renormalization uses only the algebraic structure of the
underlying theory and does not require any particular regularization. 
So it may be applied to any anomaly free model. However to provide the 
validity of STI for renormalized Green functions in this approach is a 
nontrivial problem and calculations are rather clumsy. Note that it 
refers also to the models without chiral fermions.

In the present paper I propose a new renormalization procedure which 
shares the advantages of both these approaches. The procedure is 
regularization independent and at the same time provides automatically 
STI for renormalized Green functions.

In the second section I illustrate the method by applying it to QED. 
In the third section the renormalization of the Yang-Mills theory is 
constructed. In the fourth section I discuss a simple anomaly free 
model with chiral fermions.

 \section { Invariant regularization of QED.}

In Quantum Electrodynamics formal WT identities for proper Green 
functions may be written in the form:
\begin{equation}
 \Gamma_{\mu}(p,0)=e \partial_{\mu} \Sigma(p)
 \label{1}
 \end{equation}
where $ \Gamma_{\mu}$ is the proper photon-electron vertex and 
$ \Sigma$ is the electron self energy. We shall define the 
renormalized Green functions which satisfy automatically this identity.

The renormalization procedure works loopwise. We start by defining the
renormalized one-loop electron self energy. The most general structure 
of $\Sigma$ is
\begin{equation}
 \Sigma(p)=( \hat{p}-m) \Sigma_1(p^2)+ \Sigma_2(p^2)
\label{2}
\end{equation}
Renormalized electron self energy is given by the eq.
\begin{equation}
\Sigma^r(p)=( \hat{p}-m)[ \Sigma_1(p^2)- \Sigma_1(a^2)]+ \Sigma_2(p^2)
- \Sigma_2(m^2)
\label{3}
\end{equation}
Here $a$ is some arbitrary (infrared finite) normalization point. Eq.
(\ref{3}) guarantees that the electron propagator has a pole at $p^2
=m^2$.

The renormalized vertex function is defined by the equation:
\begin{equation}
\Gamma^r_{\mu}(p,q)= \Gamma_{\mu}(p,q)- \Gamma_{\mu}(p,0)+e \partial
_{\mu} \Sigma^r(p)
\label{4}
\end{equation}
Here $\Sigma^r$ is the finite renormalized function defined by the 
eq.(\ref{3}). The vertex function diverges logarithmically, so 
$\Gamma^r_{\mu}$  is obviously finite and hence regularization 
independent. By construction it satisfies WTI:
\begin{equation}
\Gamma^r_{\mu}(p,0)=e \partial_{\mu} \Sigma^r(p)
\label{5}
\end{equation}
At first sight the subtraction (\ref{4}) may seem to be nonlocal. 
However it is not. Locality of our procedure follows from the fact 
that in any regularization scheme the difference
\begin{equation}
\Gamma_{\mu}(p,0)-e \partial_{\mu} \Sigma^r(p)
\label{6}
\end{equation}
is a local polynomial.

An alternative way to prove the locality and gauge invariance of our
renormalization procedure, which will be useful for extension to 
non-Abelian models, is the following. The definition of the 
renormalized Green functions (eqs \ref{3}, \ref{4}) is regularization 
independent. So we can consider it in some gauge invariant 
regularization scheme. Let us rewrite the eq.(\ref{4}) in the form
\begin{equation}
\Gamma^r_{\mu}(p,q)= \Gamma^{inv}_{\mu}(p,q)+(z_2-1) \gamma_{\mu}
-[ \Gamma^{inv}_{\mu}(p,0)+(z_2-1) \gamma_{\mu}-e \partial_{\mu} 
\Sigma^{inv,r}(p)]
\label{7}
\end{equation}
In an invariant regularization scheme the term in square bracketts 
vanishes by virtue of WTI, and $ \Gamma^r_{\mu}(p,q)= 
\Gamma^{inv,r}_{\mu}(p,q)$. In a gauge
invariant regularization the Green functions $\Gamma^{inv,r}_{\mu}$ 
and $\Sigma^{inv,r}$, renormalized by the counterterms obeying the 
relation $z_1=z_2$, satisfy WT identity. The locality of the procedure 
is manifest.

Gauge invariant renormalization of the vacuum polarization makes no 
problems. The polarization tensor has the structure:
\begin{equation}
\Pi_{\mu \nu}(p)=(g_{\mu \nu}p^2-p_{\mu}p_{\nu}) \Pi_1(p^2)+ 
p_{\mu}p_{\nu}\Pi_2(p^2)+g_{\mu \nu} \Pi_3(p^2)
\label{8}
\end{equation}
The renormalized polarization tensor is defined by the equation
\begin{equation}
\Pi^r_{\mu \nu}= P_{\mu \alpha}[\Pi_{\alpha \nu}(p)- \Pi_{\alpha \nu}
(0)]-p^2P_{\mu \nu} \Pi_1(b^2)
\label{9}
\end{equation}
Here $b$ is again some arbitrary normalization point and $P_{\mu \nu}
=g_{\mu \nu}-p_{\mu}p_{\nu}p^{-2}$ is the transversal projection 
operator. The function $ \Pi^r_{\mu \nu}$ is finite, hence 
regularization independent. It obviously satisfies the transversality 
condition
\begin{equation}
p_{\mu} \Pi^r_{\mu \nu}(p)=0
\label{10}
\end{equation}
required by gauge invariance.

To complete the one-loop renormalization we have to consider the 
diagram with four photon external lines. The differential WTI for 
the function $\Pi_{\mu \nu \rho \sigma}(p,q,k)$ reads:
\begin{equation}
\Pi_{\mu \nu \rho \sigma}(p,q,0)=0
\label{11}
\end{equation}
This equation is preserved in any gauge invariant regularization 
scheme. In arbitrary scheme $ \Pi_{\mu \nu \rho \sigma}(p,q,0)$ is 
a zero order polynomial (constant tensor).

We define the renormalized four-point function by the equation:
\begin{equation}
\Pi^r_{\mu \nu \rho \sigma}(p,q,k)= \Pi_{\mu \nu \rho \sigma}
(p,q,k)- \Pi_{\mu \nu \rho \sigma}(p,0,0)
\label{12}
\end{equation}
As the function $\Pi_{\mu \nu \rho \sigma}$ diverges logarithmically, 
the renormalized function defined by the eq.(\ref{12}) is obviously 
finite. It also satisfies WTI (\ref{11}). Locality of the procedure 
holds for the same reasons as above.

Renormalization of the Green functions according to eqs (\ref{3}, 
\ref{4}, \ref{9}, \ref{12}) is equivalent to introducing the 
following one-loop counterterms to the QED Lagrangian:
\begin{eqnarray}
L^r=- \frac{z_3}{4}F_{\mu \nu}F_{\mu \nu}+iz_2 \bar{\psi}(\hat
{\partial}-m) \psi+ \nonumber \\
+ \delta m \bar{\psi} \psi+ez_1 \bar{\psi} \gamma_{\mu} \psi A_
{\mu}+z_4(A_{\mu}A_{\mu})^2+ \delta \mu A_{\mu}^2
\label{13}
\end{eqnarray}
Here
\begin{eqnarray}
1-z_2= \Sigma_1(a^2); \quad \delta m=- \Sigma_2(m^2); \quad (1-z_1) 
\gamma_{\mu}=\Gamma_{\mu}(p,0)-e \partial_{\mu} \Sigma^r(p) \nonumber \\
1-z_3= \Pi_1(b^2); \quad \delta \mu=- \Pi_3(0); \quad z_4=- \frac{1}{3}
\Pi_{0000}(p,0,0)
\label{14}
\end{eqnarray}
In an invariant regularization scheme
\begin{equation}
\delta \mu=0; \quad z_4=0; \quad z_1=z_2
\label{15}
\end{equation}

Assuming that the one loop counterterms are introduced according to 
eqs(\ref{13},\ref{14}), we define the renormalized two loop functions 
by the equations(\ref{3}, \ref{4}, \ref{9}, \ref{12}). The 
divergencies in subgraphs of the diagrams corresponding to these 
functions are killed by 
the one-loop counterterms (\ref{13}), therefore they may diverge only 
superficially. Repeating literally the discussion given above, we 
prove the locality and gauge invariance of our procedure at two-loop 
level. Extension to higher loops is straightforward.

\section{Renormalization of Yang-Mills theory}.

In this section we apply the same idea to renormalization of Yang-
Mills theory. The two-point Green functions will be renormalized 
with the help of local subtractions compatible with gauge invariance, 
and renormalized vertex functions will be defined by means of nonlocal 
subtractions supplemented by addition of explicitly known finite 
terms restoring the locality and gauge invariance.

In a covariant $ \alpha$-gauge the effective action of Yang-Mills 
theory looks as follows
\begin{equation}
S= \int d^4x[- \frac{1}{4}F^a_{\mu \nu}F^a_{\mu \nu}+ \frac{1}{2 
\alpha}(\partial_{\mu}A_{\mu})^2+ 
+ \bar{c}^a \partial_{\mu}( \delta^{ab} \partial_{\mu}-gt^{abc}A^c_
{\mu})c^b]
\label{16}
\end{equation}

We want to define renormalized Green functions so that they satisfy
automatically ST identities. For the Yang-Mills self energy these 
identities reduce to the condition of transversality . So the one-
loop renormalized Yang-Mills field self energy may be defined in 
the same way as in QED, by the eq.(\ref{9}).

Next identity relates the three point function for the Yang-Mills 
field with the ghost-gauge interaction vertex and ghost and gauge 
fields propagators.

It follows from the structure of ghost interaction vertex and Lorentz
invariance, that the ghost field self energy has the form
\begin{equation}
\Pi^{ab}_G(p^2)= \delta^{ab}p^2 \Pi(p^2)
\label{17}
\end{equation}
We renormalize it by making a subtraction at arbitrary infrared finite 
point
\begin{equation}
\Pi^r(p^2)= \Pi(p^2)- \Pi(c^2)
\label{18}
\end{equation}

The ghost-gauge vertex also may be renormalized at will. It diverges
logarithmically, and it follows from the structure of interaction that 
the corresponding local structure is proportional to $t^{abc}k_{\mu}$, 
where $k_{\mu}$ is the ghost field momentum. Therefore we can perform 
the renormalization by the following prescription:
\begin{equation}
\Gamma^{abc,r}_{\mu}(k,p)= \Gamma^{abc}_{\mu}(k,p)-k_{\mu} \Gamma^
{abc}(b^2) 
\label{19}
\end{equation}
\begin{equation}
\Gamma^{abc}_{\mu}(k,p)_{p=0}=k_{\mu} \Gamma^{abc}(k^2)
\label{20}
\end{equation}
In this equation $p$ stands for the gauge field momentum.

The subtractions (\ref{9}, \ref{18}, \ref{19}) are equivalent to 
introducing to the Lagrangian the following counterterms
\begin{equation}
- \frac{z_2-1}{4}( \partial_{\mu}A^a_{\nu}-
\partial_{\nu}A^a_{\mu})(\partial_{\mu}A^a_{\nu}- \partial_{\nu}
A^a_{\mu})+(\tilde{z}_2-1) \bar{c}^a \Box c^a-( \tilde{z}_1-1)gt^
{acb} \bar{c}^a \partial_{\mu}(A^c_{\mu}c^b)
\label{21}
\end{equation}

\begin{figure}
\epsfxsize=0.3\textwidth
\centerline{\epsfbox{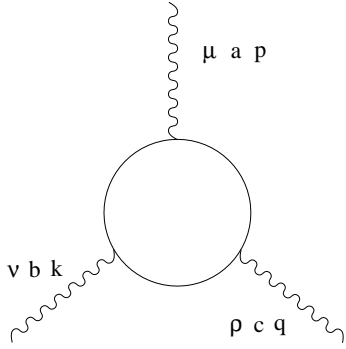}}
\caption{Three-point vertex function for Yang-Mills fields.}
\label{fig1}
\end{figure}
Now we should define the renormalized three-point Yang-Mills field 
vertex, shown at fig \ref{fig1}, in such a way it satisfies the ST 
identities. We firstly assume that some gauge invariant 
regularization scheme is introduced. Then one can get in a usual 
way the following identity for renormalized correlators
\begin{eqnarray}
\frac{i}{\alpha}<B^a_{\mu}(x)B^b_{\nu}(y) \partial_{\rho}B^c_{\rho}
(z)>^r= \tilde{z}_2< \partial_{\mu}c^a(x) \bar{c}^c(z)B^b_{\nu}(y)>
^r \nonumber\\ - \tilde{z}_2 \tilde{g}t^{ade}<B^d_{\mu}(x)c^e(x)
 \bar{c}^c(z)B^b_{\nu}(y)>^r+(x \rightarrow y, a \rightarrow b, \mu 
 \rightarrow \nu)
 \label{22}
 \end{eqnarray}
 Here the parameter $ \tilde{g}$ is the effective coupling constant 
 which enters the gauge transformations in the renormalized theory
\begin{equation}
\delta A^a_{\mu}= \partial_{\mu} \alpha^a- \tilde{g}t^{abc}A^b_{\mu}
\alpha^c
\label{23}
\end{equation}

Our definition of the renormalized three-point Yang-Mills vertex is 
based on the identity(\ref{22}). So we firstly transform the eq.
(\ref{22}) to a more convinient form. In this transformation we shall 
use the quantum equations of motoin for the ghost field $c$. These 
equations look as follows
\begin{equation}
\int[ \Box c^a(x)- \tilde{g}t^{aed} \partial_{\mu}(B^e_{\mu}(x)c^d(x))+
\tilde{z}_2^{-1} \eta^a(x)]e^{iL_R}dB_{\mu}d \bar{c}dc=0
\label{26a}
\end{equation}
Differentiating this equation with respect to $ \eta^b(y)$ and
$J^c_{\mu}(z)$ we have
\begin{eqnarray}
< \Box c^a(x) \bar{c}^b(y)>^r- \tilde{g}t^{aed}< \partial_{\mu}
(B^e_{\mu}(x)c^d(x)) \bar{c}^b(y)>^r \nonumber\\
+ \tilde{z}_2^{-1} \delta^{ab} \delta (x-y)=0
\label{27}
\end{eqnarray}
and
\begin{equation}
< \Box c^a(x) \bar{c}^c(z)B^b_{\nu}(y)>^r- 
 \tilde{g}t^{aed}< \partial_{\mu}(B^e_{\mu}(x)c^d(x)) \bar{c}^c(z)
B^b_{\nu}(y)>^r=0
\label{28}
\end{equation}
Note that eqs(\ref{26a}, \ref{27}, \ref{28}) are valid also in the 
"partially renormalized" theory, described by the Lagrangian, including 
the counterterms (\ref{21}) (no counterterms for three- and four-point 
gauge field vertices). It follows from the eq.(\ref{28}) that the sum 
of the first two terms in the l.h.s. of identity (\ref{22}) is 
transversal with respect to differentiation over $x_{\mu}$ and remaining 
terms are transversal with respect to differentiation over $y_{\nu}$. 
Using this observation we can rewrite the identity (\ref{22}) as follows. 
We firstly perform a Fourier transformation of eq.(\ref{22}). Then 
cutting the Yang-Mills field propagators 
corresponding to  the external lines of the three-point correlator at the 
l.h.s. of the eq.(\ref{22}) and taking into account the transversality of 
Yang-Mills field self energy, we can write for the proper vertex function 
$ \Gamma^{abc,r}_{\mu \nu \rho}(p,q)$ the following identity:
\begin{eqnarray}
q_{\rho} \Gamma^{abc,r}_{\mu \nu \rho}(p,q)-q^2[(G^{-1}_{tr})^r_{\mu
\alpha}(p)G^r(q) \Gamma^{abc,r}_{\alpha \nu}(p,q)+\nonumber\\
+( \mu \rightarrow \nu, a \rightarrow b, p \rightarrow -p-q)]=0
\label{36}
\end{eqnarray}
In deriving this equation we used the transversality of the r.h.s. of
eq.(\ref{22}) discussed above,  and replaced the inverse gauge field
propagator $G_{\mu \nu}$ by its transversal part. After such projection 
the first term in the l.h.s. of eq.(\ref{22}), which is longitudinal, 
does not contribute. In this equation
\begin{eqnarray}
\delta^{ab}(G^{-1}_{tr})^r_{\mu \nu}=P_{\mu \alpha}(G^{-1})^{ab,r}_{\alpha
\nu}\nonumber\\
 \delta^{ab}G^r(p)=G^{ab,r}(p)
\label{25}
\end{eqnarray}
$G^{ab,r}(p)$ is the renormalized ghost field propagator.
The function $\Gamma^{abc,r}_{\mu \nu}(p,q)$ is the Fourier transform of
the expectation value of the composite operator
\begin{equation}
\tilde{\Gamma}^{abc,r}_{\mu \nu}=(G^{-1})^r_{\nu \alpha}(y)(G^{-1})^r(z)<
\tilde{z}_2 \tilde{g}t^{aed}B^e_{\mu}(x)c^d(x) \bar{c}^c(z)B^b_{\nu}(y)>^r
\label{25a}
\end{equation}
Differentiating the equality (\ref{36}) with respect to $q_{\rho}$ and 
putting $q=0$ we get the differential identity, which will be used for 
the definition of renormalized three-point Yang-Mills field vertex
\begin{eqnarray}
\Gamma^{abc,r}_{\mu \nu \rho}(p,0)- \frac{\partial}{ \partial q_{\rho}}
[q^2(G^{-1}_{tr})^r_{\mu \alpha}(p)G^r(q) \Gamma^{abc,r}_{\alpha \nu}(p,q)
+\nonumber\\ +( \mu \rightarrow \nu, a \rightarrow b, p \rightarrow -p-q)]
_{q=0}=0
\label{37}
\end{eqnarray}
The renormalized proper vertex function for gauge fields is defined by 
the equation
\begin{eqnarray}
\Gamma^{abc,r}_{\mu \nu \rho}(p,q)= \Gamma^{abc}_{\mu \nu \rho}(p,q)- 
\{\Gamma^{abc}_{\mu \nu \rho}(p,0)+\Gamma^{abc}_{\mu \nu \rho}(0,q)- 
\nonumber\\
 \frac{ \partial}{ \partial q^{\rho}}[q^2(G^{-1}_{tr})^r_{\mu
\alpha}(p)G^r(q) \Gamma^{abc,r}_{\alpha \nu}(p,q)+ 
(\nu \rightarrow \mu, a \rightarrow b,
p \rightarrow -p-q)]_{q=0}- \nonumber\\ 
\frac{ \partial}{ \partial p^{\mu}}[p^2(G^{-1}_{tr})^r_{\rho \beta}(q)G^r
(p) \Gamma^{cba,r}_{\beta \nu}(q,p)+(\nu \rightarrow \rho, c \rightarrow 
b, q \rightarrow -p-q)]_{p=0} \}
\label{24}
\end{eqnarray}

The divergent part of $ \Gamma^{abc}_{\mu \nu \rho}(p,q)$ is the first 
order polynomial in $p,q$. Therefore the sum of the three first terms in 
the r.h.s. of eq.(\ref{24}) is finite. Below we shall prove that the 
remaining terms in the r.h.s. of eq.(\ref{24}) are also finite.

As it was mentioned above the r.h.s. of eq.(\ref{24}) includes the 
expectation value of the composite operator $ \Gamma^{abc}_{\mu \nu}$. 
So we have to prove that the counterterms (\ref{21}) make it finite. In 
this proof we again shall use the quantum equations of motion for gauge 
fields (\ref{26a}, \ref{27}, \ref{28}).

\begin{figure}
\epsfxsize=0.5\textwidth
\centerline{\epsfbox{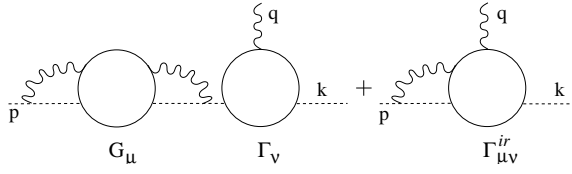}}
\caption{Decomposition of $ \Gamma_{\mu \nu}$ into one-particle reducible and
irreducible parts.}
\label{fig2}
\end{figure}

The function $\tilde{\Gamma}_{\mu \nu}$ may be separated into one particle
reducible and irreducible parts as follows (see fig \ref{fig2}):
\begin{equation}
\tilde{\Gamma}^{abc,r}_{\mu \nu}(x,y,z)= \int \tilde{G}^{ad}_{\mu}(x-x')
\tilde{\Gamma}^{dbc,r}_{\nu}(x',y,z)dx'
+( \tilde{\Gamma}^{ir})^{abc}_{\mu \nu}(x,y,z)
\label{29}
\end{equation}
Here $ \tilde{\Gamma}^{ir}_{\mu \nu}$ denotes the strongly connected part 
of $ \tilde{\Gamma}_{\mu \nu}$, and $ \Gamma^r_{\nu}$ is the renormalized
ghost-gauge field vertex. Finally
\begin{equation}
\tilde{G}^{ab}_{\mu}(x-x')=< \tilde{z}_2 \tilde{g}t^{aed}B^e_{\mu}(x)c^d(x)
\bar{c}^b(x')> 
\label{30}
\end{equation}
By virtue of Lorentz invariance $G^{ab}_{\mu}(p)$ is proportional to
$p_{\mu}$. Then the eq.(\ref{27}) gives
\begin{equation}
G^{ab}_{\mu}(p)= \tilde{z}_2p_{\mu}G^{ab}(p)+\delta^{ab}p_{\mu}p^{-2}
\label{32}
\end{equation}
Substituting the decomposition (\ref{29}, \ref{32}) into eq.(\ref{28}) we
get
\begin{equation} 
\Gamma^{abc,r}_{\nu}(p,q)+ ip_{\mu} (\Gamma^{ir})^{abc}_{\mu \nu}(p,q)=0
\label{33}
\end{equation}
This equation shows that the function $p_{\mu}(\Gamma^{ir})
^{abc}_{\mu \nu}$ is finite. As the corresponding integral diverges 
logarithmically, Lorentz invariance implies that $( \Gamma^{ir})^{abc}_{\mu 
\nu}(p,q)$ is also finite.

The function $\Gamma^{abc}_{\mu \nu}(p,q)$ enters into eq.(\ref{24}) being
multiplied by the transverse projector. As the eqs (\ref{29}-\ref{32})
show the reducible part of $\Gamma_{\mu \nu}$ is longitudinal and hence
only irreducible part of $ \Gamma_{\mu \nu}$ contributes to eq.(\ref{24}).
It proves the finiteness of the r.h.s. of eq.(\ref{24}).

Therefore the eq.(\ref{24}) indeed defines the finite function
$\Gamma^{abc,r}_{\mu \nu \rho}(p,q)$ which does not depend on a particular 
regularization scheme used for its calculation.

Now we shall prove that the definition (\ref{24}) corresponds to 
subtraction from $ \Gamma^{abc}_{\mu \nu \rho}(p,q)$ a local polynomial 
and that $ \Gamma^{abc,r}_{\mu \nu \rho}(p,q)$ satisfies ST identities.

As the eq.(\ref{24}) is regularization independent we may use for the proof
a particular regularization for which we choose some gauge invariant
regularization. With an invariant regularization all one loop divergencies
may be eliminated by introduction to the effective Lagrangian the
counterterms satisfying the relations
\begin{equation}
z_1z_2^{-1}= \tilde{z}_1 \tilde{z}_2^{-1}; \quad z_4=z_1^2z_2^{-1}
\label{34}
\end{equation}
where the counterterms $z_1$ and $z_4$ renormalize the three and four-point
gauge fields vertices respectively:
\begin{eqnarray}
\Gamma^{abc,r}_{\mu \nu \rho}(p,k,q)= \Gamma^{abc}_{\mu \nu
\rho}(p,k,q)+\nonumber\\
i(z_1-1)t^{abc}[(p-k)_{\rho}g_{\mu \nu}+(k-q)_{\mu}g_{\nu
\rho}+(q-p)_{\nu}g_{\mu \rho}]
\label{35}
\end{eqnarray}

Let us rewrite the sum of the first three terms in the r.h.s. of
eq.(\ref{24}) as follows
\begin{eqnarray}
( \Gamma^{abc}_{\mu \nu \rho}(p,q)+i(z_1-1)t^{abc}[(p-k)_{\rho}g_{\mu
\nu}+(k-q)_{\mu}g_{\nu \rho}+(q-p)_{\nu}g_{\mu \rho}])-\nonumber\\
-( \Gamma^{abc}_{\mu \nu \rho}(p,0)+i(z_1-1)t^{abc}[2p_{\rho}g_{\mu
\nu}-p_{\mu}g_{\nu \rho}-p_{\nu}g_{\mu \rho}])-\nonumber\\
-( \Gamma^{abc}_{\mu \nu \rho}(0,q)+i(z_1-1)t^{abc}[q_{\rho}g_{\mu
\nu}-2q_{\mu}g_{\nu \rho}+q_{\nu}g_{\mu \rho}])
\label{35a}
\end{eqnarray}
Here $ k=-p-q; \quad z_1=( \tilde{z}_1 \tilde{z}_2^{-1})z_2$. According 
to eq.(\ref{35}) the first line is this equation is the three-point gauge 
field vertex renormalized in a gauge invariant way. Analogously, the second 
and third lines represent the functions $ \Gamma^{abc,r}_{\mu \nu \rho}
(p,0)$ and $\Gamma^{abc}_{\mu \nu \rho}(0,q)$ renormalized in a gauge 
invariant way. These functions satisfy the renormalized identity (\ref
{37}). Comparing these identities with the r.h.s. of eq.(\ref{24}) we see 
that the terms in curly bracketts cancel, and the function $ \Gamma^{abc,r}
_{\mu \nu \rho}$ coincides with the function (\ref{35}), obtained in a 
gauge invariant regularization scheme by means of a local subtraction. This
function obviously satisfies ST identities.

The last one-loop diagram to be analyzed is the proper four-point
Yang-Mills vertex. In any regularization this function has the following
structure
\begin{equation}
\Pi^{abcd}_{\mu \nu \rho \sigma}(p,q,k)=P[At^{abe}t^{cde}g_{\mu
\rho}g_{\nu \sigma}+B \delta^{ab} \delta^{cd}g_{\mu \nu}g_{\rho
\sigma}]+O(p,q,k)
\label{38}
\end{equation}
Here $P$ is the simmetrization operator with respect to the pairs $(a,
\mu;b, \nu; c, \rho; d, \sigma)$, $A$ and $B$ are some (diverging when
regularization is removed) constants, $O(p,q,k)$ is a finite function.

It follows from equation (\ref{38}) that 
\begin{equation}
B= \frac{1}{3}p^{-4}p_{\mu}p_{\nu}p_{\rho}p_{\sigma}
\Pi^{aaaa}_{\mu \nu \rho \sigma}(p,0,0)+O_1
\label{39}
\end{equation}
(no summation over $a$)
\begin{equation}
A= \frac{1}{3Np^2}P_{\nu \rho}(p)p_{\mu}p_{\sigma} \Pi^{abab}_{\mu \nu
\rho \sigma}(p,0,0)+O_2
\label{40}
\end{equation}
(no summation over $a$).
Here $N=\sum_{b,e}t^{abe}t^{abe}$, $P_{\mu \nu}$ is the
transversal projector operator and $O_1, O_2$ are some finite functions.

\begin{figure}
\epsfxsize=0.3\textwidth
\centerline{\epsfbox{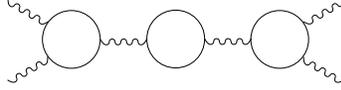}}
\caption{Weakly connected diagrams contributing to the l.h.s. of eq(\ref{41}).}
\label{fig3}
\end{figure}

We define the renormalized four-point function as follows:
\begin{eqnarray}
\Pi^{abcd,r}_{\mu \nu \rho \sigma}(p,q,k)=\Pi^{abcd}_{\mu \nu \rho \sigma}
(p,q,k)- \nonumber\\ 
P(t^{abe}t^{cde}g_{\mu \rho}g_{\nu \sigma}) 
(3Np^2)^{-1}P_{\beta \gamma}(p)p^{\alpha}p^{\delta}
(\Pi_{con})^{mnmn}_{\alpha \beta \gamma \delta}(p,0,0)-\nonumber\\
P(g_{\mu \nu}g_{\sigma \rho} \delta^{ab}
\delta^{cd})3^{-1}p^{-4}p^{\alpha}p^{\beta}p^{\gamma}p^{\delta}
(\Pi_{con})^{mnmn}_{\alpha \beta \gamma \delta}(p,0,0)
\label{41}
\end{eqnarray}
Note that the subtracted terms at the r.h.s. of this equation include the
functions $\Pi_{con}(p,0,0)$ which represent all connected one loop 
diagrams with four external gauge field lines. They include apart from the 
proper four-point vertex also weakly connected diagrams shown at fig \ref
{fig3}. The divergent subgraphs in these diagrams are assumed to be 
renormalized as described above.

By virtue of eqs (\ref{39}, \ref{40}) the renormalized function $
\Pi^{abcd,r}_{\mu \nu \rho \sigma}(p,q,k)$ defined by the eq.(\ref{41}) 
is finite and hence regularization independent.

To prove the locality and gauge invariance we introduce again some
invariant regularization. Then STI for the four point function imply the
relation
\begin{eqnarray}
< \partial_{\mu}B^a_{\mu}(x) \partial_{\nu}B^b_{\nu}(y)
\partial_{\rho}B^c_{\rho}(z) \partial_{\sigma}B^d_{\sigma}(u)>= \nonumber\\
= \delta(x-u) \delta(y-z) \delta^{ad} \delta^{bc}+(x,a \rightarrow y,b)+
(x,a \rightarrow z,c)
\label{42}
\end{eqnarray}
This equation shows that the connected part of the correlator (\ref{42}) 
is equal to zero. In renormalized theory this identity holds provided the
counterterms satisfy the equation
\begin{equation}
\tilde{z}_4=0; \quad z_4=z_1^2z_2^{-1}
\label{44}
\end{equation}
Therefore
\begin{equation}
p_{\mu}p_{\sigma} (\Pi^r_{con})^{abcd}_{\mu \nu \rho \sigma}(p,0,0)=0
\label{43}
\end{equation}

Let us replace all the functions $ \Pi^{abcd}_{\mu \nu \rho \sigma}$ at 
the r.h.s. of eq.(\ref{41}) by invariantly renormalized functions $(
\Pi^r_{inv})^{abcd}_{\mu \nu \rho \sigma}$
\begin{equation}
(\Pi^r_{inv})^{abcd}_{\mu \nu \rho \sigma}=( \Pi_{inv})^{abcd}_
{\mu \nu \rho \sigma}+(z_4-1)P(t^{abe}t^{cde}g_{\mu \nu}g_{\rho \sigma})
\label{45a}
\end{equation}
Such a substitution does not change the eq.(\ref{41}) as the terms $ \sim
(z_4-1)$ in the first and the second lines cancel. Invariantly 
renormalized vertices satisfy the identity (\ref{43}). Therefore the 
terms in the second and third lines of eq.(\ref{41}) vanish.

 So we conclude that in a gauge invariant scheme
\begin{equation}
\Pi^{abcd,r}_{\mu \nu \rho \sigma}= (\Pi^{inv})^{abcd}_{\mu \nu \rho
\sigma}(p,q,k)+(z_4-1)P(t^{abe}t^{cde}g_{\mu \rho}g_{\nu \sigma})
\label{46}\end{equation}
As was discussed above the l.h.s. of eq.(\ref{46}) does not depend on 
a regularization. Therefore in any scheme our renormalized function 
coincides with the function which was obtained by means of a local 
subtraction and satisfies ST identities. It completes the 
renormalization of one loop-diagrams.

Extension to the higher loops is done exactly as in the case of QED.
Introducing the one-loop counterterms we are left with the two-loop
diagrams which diverge only superficially. We can define the 
renormalized two-loop correlation functions by the same equations as 
above. They are obviously finite and regularization independent. The 
proof of locality is identical to the one loop-case.

\section{A model with chiral fermions}

As the last example we consider a simple model which includes chiral
fermions. We choose the Abelian sector of Salam-Weinberg model with 
the Higgs interaction switched off. This model is known to be anomaly 
free but not a vectorlike one

The Lagrangian looks as follows:
\begin{eqnarray}
L=- \frac{1}{4}( \partial_{\mu}B_{\nu}- \partial_{\nu}B_{\mu})^2+i 
\sum_{q \pm} \bar{\psi}_{q \pm} \gamma_{\mu}( \partial_{\mu}-ig_{q 
\pm}B_{\mu})\psi_{q \pm}+\nonumber\\ +i \sum_{l \pm} \bar{\psi}_{l 
\pm} \gamma_{\mu}(\partial_{\mu}-ig_{l \pm}B_{\mu}) \psi_{l \pm}
\label{47}
\end{eqnarray}
Here $ \psi_{q \pm}$ and $\psi_{l \pm}$ represent left (right) handed
fields of quarks and leptons respectively. The corresponding charges 
are denoted by $g_{q \pm}$ and $g_{l \pm}$. Due to the condition
\begin{equation}
\sum_{q \pm}g^3_{q \pm}+ \sum_{l \pm}g^3_{l \pm}=0
\label{48}
\end{equation}
the triangle anomaly is absent.

Renormalized gauge field and fermion propagators are defined by the
eqs( \ref{3}, \ref{9}) of the section 2. The renormalized gauge
field-fermion vertices are defined as in eq.(\ref{4}).
\begin{equation}
\Gamma^{\mu,r}_{q \pm,l \pm}= \Gamma^{\mu}_{q \pm,l \pm}(p,q)-
\Gamma^{\mu}_{q \pm,l \pm}(p,0)+g_{q \pm,l \pm} \partial_{\mu} 
\Sigma^r_{q \pm,l \pm}(p)
\label{49}
\end{equation}

The four-point gauge field correlator can be renormalized according to
eq.(\ref{12}), but in this case the subtracted term has to include all
connected diagrams with four external gauge field lines.

The only essentially new moment is a possible renormalization of the
three-point gauge field vertex. In QED it is zero due to Furry theorem, 
but in our case it does not vanish.

WTI for this function have the same form as for the four-point 
function
\begin{equation}
p_{\mu} \Gamma_{\mu \nu \rho}(p,q)=0; \quad \Gamma_{\mu \nu \rho}(0,q)
=0
\label{50}
\end{equation}
Accordingly we choose the following definition of renormalized vertex
function
\begin{equation}
\Gamma^r_{\mu \nu \rho}(p,q)= \Gamma_{\mu \nu \rho}(p,q)-
\Gamma_{\mu \nu \rho}(p,0)-\Gamma_{\mu \nu \rho}(0,q)
\label{51}
\end{equation}

This function is finite and can be calculated in any regularization 
scheme. In a gauge invariant scheme one has due to WTI(\ref{50})
\begin{equation}
 \Gamma^r_{\mu \nu \rho}(p,q)= \Gamma^{inv}_{\mu \nu \rho}(p,q)
\label{53}
\end{equation}

 \section{Discussion}

The renormalization procedure presented above combines the advantages 
of algebraic renormalization and invariant regularization schemes. It 
is regularization independent and may be applied to any anomaly free 
model. At the same time it preserves gauge invariance at all stages 
and does not require additional fine tuning of counterterms to restore 
the symmetry. It also avoids the problem of infrared singularities in 
renormalization of Green functions. From the point of view of practical 
calculations it may be more complicated than dimensional regularization 
if one is interested only in some particular diagram. Our procedure is 
recurrent and it requires knowleadge of lower order diagrams for some 
particular configurations of external momenta. However it seems well 
suited for systematic calculation of Feynman diagrams up to a given 
order. It is important to emphasise that our renormalization procedure 
may be applied directly to the models where dimensional regularization 
fails, in particular to the Standard Model and supersymmetric theories.
It is worth to mention that to make practical calculations easier one
can combine our renormalization procedure with the dimensional 
regularization. For example in chiral fermion models one may define the 
$ \gamma_5$ matrix as the product $ \gamma_5= \gamma_0 \gamma_1 
\gamma_2 \gamma_3$ (\cite{H.V},\cite{B.M}), and then apply the dimensional 
regularization. For diagrams with virtual fermion lines this definition 
is known to break chiral invariance. However the symmetry may be easily 
restored with the help of subtractions described above.

 {\bf Acknowledgements.} \\ This work was done while the author was
 visiting Max-Planck-Institute for Physics in Munich. I wish to thank
 D.Maison and J.Wess for hospitality and Humboldt Foundation for a 
 generous support. My thanks to the members of Theoretical group for 
 helpful discussion. This researsh was supported in part by Russian 
 Basic Research Fund under grant 99-01-00190 and by the president 
 grant for support of leading scientific schools 0015-96046.$$ ~ $$ 

\begin{thebibliography}{99}
{\small \bibitem{W}J.C.Ward, Phys.Rev. 77 (1950) 2931.
 \bibitem{T} J.Takahashi, Nuovo Cim. 6 (1957) 370.
\bibitem{Sl} A.A.Slavnov, Theor.Math.Phys. 10 (1972) 99.
\bibitem{Tl} J.G.Taylor, Nucl.Phys. B33 (1971) 436.
\bibitem{H.V} G.'tHooft, M.Veltman, Nucl.Phys. B44 (1972) 189.
\bibitem{B.M} P.Breitenlohner, D.Maison, Comm.Math.Phys. 52 (1977) 11.
 \bibitem{Sl2} A.A.Slavnov, Theor.Math.Phys. 13 (1972) 174.
\bibitem{L.Z} B.W.Lee, J.Zinn-Justin, Phys.Rev. D5 (1972) 3137.
\bibitem{B.S} T.D.Bakeyev, A.A.Slavnov, Mod.Phys.Lett. A11 (1996) 1539.
\bibitem{Wil} K.Wilson, Phys.Rev. D14 (1974) 2445.
\bibitem{B.P} N.N.Bogoliubov, O.S.Parasiuk, Sov.Phys.Dokladi 100 (1955) 25.
\bibitem{Z} W.Zimmerman, Comm.Math.Phys. 39 (1974) 81.
\bibitem{BRS} C.Becci, A.Rouet, R.Stora, Comm.Math.Phys.
42 (1975) 127.
\bibitem{P.R} O.Piguet, A.Rouet, Phys.Rep. 76 (1981).
\bibitem{Gr} P.A.Grassi, T.Hurth, M.Steinhauser, Ann.Phys. 288 (2001) 197.
\bibitem{F.S} S.A.Frolov, A.A.Slavnov, Phys.Lett. B309 (1993) 344.
\bibitem{D.S} M.M.Deminov, A.A.Slavnov, Phys.Lett. B50 (2001) 297.}
 \end {thebibliography}
  \end{document}